\documentclass{ws-procs975x65}

\begin{document}

\title{Single Wall Carbon nanotube weak links}

\author{K. Grove-Rasmussen$\dagger$ and H.~ I. ~J\O rgensen$\dagger$ and P.~E. Lindelof}

\address{Nano-Science Center\\
Niels Bohr Institute\\
Universitetsparken 5, \\
2100 Copenhagen O, Denmark\\
E-mail: k\_grove@fys.ku.dk\\
$\dagger$ These authors contributed equally to this work.}

\maketitle

\abstracts{We have reproducibly contacted gated single wall carbon
nanotubes (SWCNT) to superconducting leads based on niobium. The
devices are identified to belong to two transparency regimes: The
Coulomb blockade and the Kondo regime. Clear signature of the
superconducting leads is observed in both regimes and in the Kondo
regime a narrow zero bias peak interpreted as a proximity induced
supercurrent persist in Coulomb blockade diamonds with Kondo
resonances.}

\section{Introduction}
Carbon nanotubes have been under intense investigation for more than
a decade due to their unique mechanical and electrical properties.
Single wall carbon nanotubes (SWCNT) are truly one-dimensional
systems and in contrast to the semi-1D channels defined in a
two-dimensional electron gas they can easily be contacted to
materials with interesting properties such as superconductors (S).

A normal region between two superconductors acts as a weak
link\cite{Likharev} and effects as proximity induced supercurrent
and sub-gap structure due multiple Andreev reflections (MAR) can be
seen. In small low capacitance junctions ({\em e.g.} SWCNT) the
above effects are modified by size and charge quantization. Size
quantization leads to a discrete density of states in the normal
region with levels of width $\Gamma$ separated by $\Delta E$. The
supercurrent will be maximum in resonance (aligned with a level) and
decreases to a minimum off resonance.\cite{Beenakker,Jarillo,HIJ}
For weakly coupled SWCNTs, charge quantization gives rise to Coulomb
blockade\cite{Bockrath} which generally suppresses the
supercurrent.\cite{Glazman,Beenakker} However, when the number of
electrons on the dot is odd (net spin 1/2 on the dot) and relative
good coupling is achieved, the Kondo effect becomes important. The
Kondo effect effectively sets up a resonance with width $\sim k_B
T_K$ at the Fermi energy of the leads and in some sense turns off
Coulomb blockade. Thus a resonance exists despite Coulomb blockade
to carry the supercurrent ($k_B T_K\gtrsim\Delta$). The interplay
between Kondo/Coulomb blockade and superconductivity is under
intense interested.\cite{KGR,Glazman,Choi,Siano,BuitelaarKondo}

We here present measurements on SWCNT devices with Nb contacts
showing high quality SWCNT quantum dots and address the topic of the
supercurrent carried through Kondo resonances. From a technical
point of view it is not the first time niobium has been used as
superconducting contacts for SWCNT.\cite{Morpurgo,Haruyama} However,
to the best of our knowledge niobium based devices have not been
presented clearly revealing Kondo resonances and Coulomb blockade
before. Several groups have made week links based on carbon
nanotubes with other superconductors such as tantalum,
rhenium\cite{Kasumov} and
aluminum.\cite{Jarillo,HIJ,KGR,BuitelaarKondo,Haruyama,BuitelaarMAR}
The advantage of using niobium is its high critical temperature
$\sim 9$\,K, which should allow for measurements above 4\,K (see
below). Furthermore, a high critical temperature gives a high
superconducting energy gap $\Delta$ which increases the critical
current\cite{Beenakker} as $I_c\sim \Delta$.

\section{Sample processing}\label{subsec:prod}
The two terminal SWCNT devices presented in this proceeding is made
with two types of superconducting contacts based on niobium:
Pd/Nb/Pd and Ti/Nb/Ti. The lower metallic layer (Ti or Pd) ensures
relatively good contact to the SWCNT, while Nb has a very high
critical temperature $\sim 9$\,K. The top layer protects the Nb from
oxidation. However, we experienced that the superconducting
proximity effect from the Nb layer into the lower Ti or Pd layer is
very weak resulting in relative low critical temperatures of
$T_c\sim 2$\,K. The Nb film still have high critical temperature
$\sim 9$\,K measured on a four terminal device.
\begin{figure}[t!]
\centerline{\epsfxsize=4.1in\epsfbox{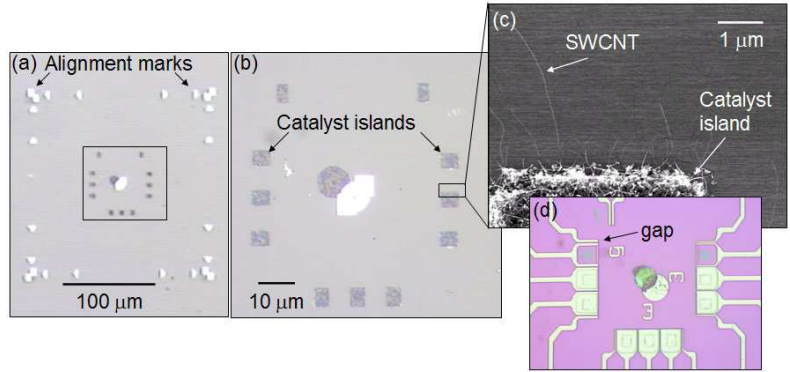}} \caption{(a) Top
view of the alignment marks used to align the EBL mask for the
catalyst islands and electrodes. In the black rectangle, catalyst
islands have subsequently been defined. (b) Zoom-in on the catalyst
islands in (a). (c) SEM Micrograph of a catalyst island after
CVD-growth. Only one or a few SWCNTs reach further than 1\,$\mu$m
from the catalyst island which makes it likely to have only one
SWCNT between the electrodes to be defined at some distance from the
catalyst island. (d) Electrodes are defined with gap sizes of 300 to
700\,nm (here 300 and 500\,nm). \label{Fig:Sampleproc}}
\end{figure}
The substrate consist of 500\,nm SiO$_2$ on a highly doped silicon
wafer (acceptors Sb, resistivity $\rho< 1-3$\,m$\Omega$\,cm), which
is used to electrostatically change the potential of the SWCNT.
First alignment marks are defined by electron beam lithography (EBL)
as shown in Fig.\ \ref{Fig:Sampleproc}a. The resist used is double
layered consisting of 6\% copolymer followed by 2\% PMMA. Both
layers are spun at 4000\,rpm, 45\,s and baked on a hotplate at
185\,$^\circ$C for 90\,s. The EBL is done on a JEOL JSM-6320F
scanning electron microscope (SEM) with Elphy software at 30\,k$e$V
with a sensitivity of 200\,$\mu$C/cm$^2$ and currents of $\sim
40$\,pA and $\sim 150$\,pA for small and big patterns, respectively.
The pattern is developed in MIBK:IPA (1:3) for 60\,s stopped by
30\,s in IPA. The sample is ashed for 20-40\,s in an oxygen plasma
prior to deposition of 60-70\,nm Cr. Lift-off is done in acetone.

The same EBL procedure is used to make the pattern for the catalyst
islands. The catalyst liquid is spun on the sample at 4000\,rpm for
150\,s and baked at 185\,$^\circ$C for 3 min.\ to ensure that the
catalyst sticks to the substrate. Lift-off is made in acetone and
ultrasound if necessary. Figure \ref{Fig:Sampleproc}b shows an
optical image of the catalyst islands. The sample is now transferred
to a chemical vapor deposition (CVD) tube furnace, where SWCNTs are
grown in a controlled mixture of argon, hydrogen and methane (Ar:
$1$\,L/min, H$_2$: $0.1$\,L/min, CH$_4$: 0.5\,L/min). The optimal
growth temperature depends on the thickness and the area of the
catalyst islands. Typically, the temperature used is in the range of
$850-960$\,$^\circ$C. After each growth the sample is examined in a
SEM on a test area, which might appear similar to the picture shown
in Fig.\ \ref{Fig:Sampleproc}c. Sometimes several trials are needed
changing the temperature to achieve an appropriate density of tubes.
The test area ensures that the SWCNTs in the regions for devices are
not exposed to the electron beam, which might damage the SWCNT. On
the catalyst island and within a micron, the nanotubes grow rather
densely. However, one or a few tubes "escape" to distances of
several microns. The contacts are aligned relative to the catalyst
island and placed some microns away by EBL (no oxygen plasma step).
Thus it is likely that only one SWCNT lies in the gap between the
two electrodes. The Pd or Ti is evaporated by thermal evaporation,
while Nb is DC-sputtered in an Ar atmosphere. Typical layers used
are 4\,nm Ti (or Pd) followed by 60\,nm Nb and capped by 10\,nm Ti
(or Pd). Finally bonding pads of Au/Cr are defined by optical
lithography.

\section{Coulomb blockade regime} We first present measurements of
a Pd/Nb/Pd contacted SWCNT with poor transparency $G\ll e^2/h$ that
leads to Coulomb blockade at low temperatures.
\begin{figure}[t!]
\centerline{\epsfxsize=4.1in\epsfbox{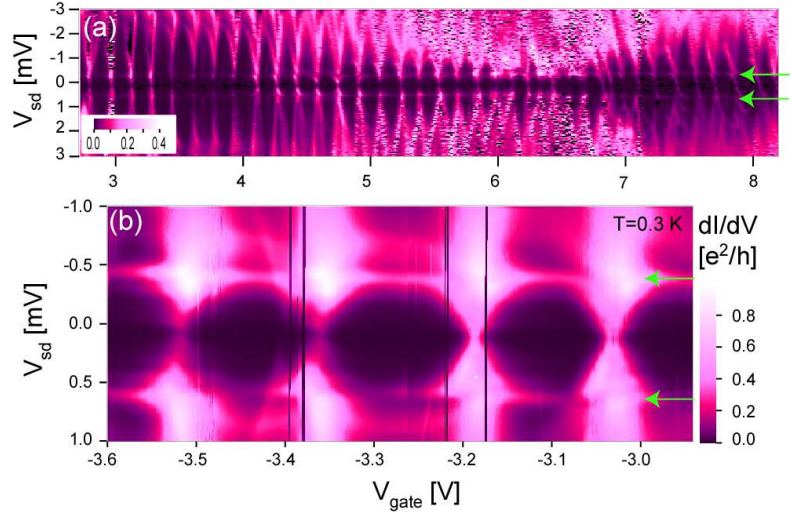}} \caption{(a) Bias
spectroscopy plot of a Pd/Nb/Pd contacted SWCNT showing regular
Coulomb blockade diamonds at $T= 0.3$\,K. Gate independent peaks in
dI/dV at low biases (green arrows) are seen throughout all diamonds
which are due to the onset of quasiparticle tunneling. The color
scale is dI/dV in $e^2/h$. (b) The onset of quasiparticle tunneling
(green arrows) is more clearly revealed in a more conducting region
corresponding to $eV_{sd}=2\Delta\sim 0.5$\,mV. Close to resonance
conductance below the gap is due to Andreev reflection. The black
vertical lines are due to noise and some gate switching is seen as
well. \label{Fig:CBbiasspecs}}
\end{figure}
Figure \ref{Fig:CBbiasspecs}a shows a bias spectroscopy plot of such
a device where clear Coulomb blockade diamonds are seen with
charging energies of $U_c\sim 2-3$\,m$e$V. At low bias a dip in the
differential conductance is observed due to the tunnel-like behavior
of the device reflecting the density of states of the
superconducting leads.\cite{Drndic} The green arrows point to the
onset of quasiparticle tunneling $eV_{sd}=2\Delta\sim \pm 500$\,$\mu
e$V. In Fig.\ \ref{Fig:CBbiasspecs}b a zoom of another slightly more
conducting gate region of the device is shown. The onset of
quasiparticle tunneling is clearly revealed (green arrows) and some
conductance in the resonances is seen below the gap which is
attributed to Andreev reflections. Such regular behavior has been
observed in several devices with niobium based contacts and poor
transparency. The onset of quasiparticle tunneling corresponds to a
critical temperature of the $T_c\sim 1.5$\,K consistent with the
features vanishing above $T_c$. This critical temperature is much
lower than the critical temperature measured in the niobium film and
is properly due to poor proximity effect into the lower palladium
layer in contact with the nanotube.

\section{Supercurrent and the Kondo effect} The next device is more
conducting and thus belong to a transparency regime where the Kondo
effect can be observed, when the SWCNT quantum dot has a net
spin.\cite{Nygaard} The contacts are made of a Ti/Nb/Ti trilayer.
\begin{figure}[t!]
\centerline{\epsfxsize=4.1in\epsfbox{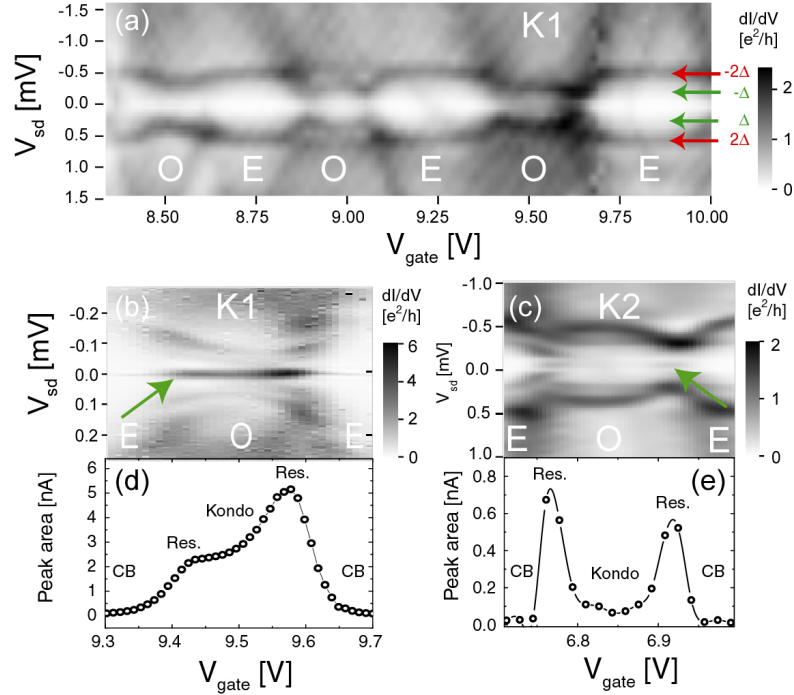}} \caption{(a) Bias
spectroscopy plot of three Kondo resonances in a Ti/Nb/Ti sample in
the superconducting state. O/E denote odd/even electron filling on
the dot. Conductance peaks at $V_{sd}=\pm 2\Delta/e$ are most
pronounced for even filling (red arrows) while $\pm \Delta/e$
conductance peaks are more pronounced for odd filling (green arrows)
due to the Kondo resonances (see text). (c) Detailed plot of Kondo
resonance K1 from (a). The green arrow points to a zero bias peak
interpreted as a noise smeared supercurrent present throughout the
gate region shown. (c) Another Kondo resonance with a similar less
pronounced zero bias peak. (d) Zero bias peak area versus gate
voltage extracted from (b). The zero bias peak is suppressed in
Coulomb blockade regions, while it exists in both Coulomb blockade-
and Kondo-resonances. (e) Similar plot as (d). The zero bias peak
area has maximum in the resonances and is also finite in the Kondo
resonance, while being suppressed in Coulomb blockade regions. Note,
that the data of (a) and (c) have been interpolated and smoothened.
\label{Fig:Kondos}}
\end{figure}
Figure \ref{Fig:Kondos}a shows a bias spectroscopy plot in a gate
region with an even-odd filling of single particle levels marked by
the letters E/O. The Kondo resonances are identified by the clear
change in magnitude of the features for the onset of quasiparticle
tunneling ($eV_{sd}=\pm 2\Delta$) and one Andreev reflection
($eV_{sd}=\pm \Delta$). For even electron filling, the onset of
quasiparticle tunneling is more pronounced because quasiparticle
tunneling is a lower order process than one Andreev reflection and
transport happens via cotunneling in the Coulomb blockade region.
However, when the Kondo effect is present (odd filling), a resonance
exists at the Fermi energy of each contact leading to resonances for
one Andreev reflection processes at $eV_{sd}=\pm \Delta$. This
qualitative explanation depends on the relative magnitude of $k_B
T_K$ and $\Delta$. We also note that this effect has been observed
in other Kondo devices.\cite{KGR,Sand}

Figure \ref{Fig:Kondos}b shows the Kondo resonance K1 at low bias
voltages where the high conductive black region (green arrow)
reflects a zero bias peak in dI/dV versus bias.\cite{KGR,HIJ} We
interpret this peak as a noise smear proximity induced supercurrent
with an estimated magnitude given by the area of the peak (not the
whole area under the peak). This interpretation is supported by the
successful analysis of an analogous zero bias peak interpreted as a
supercurrent in an open quantum dot.\cite{HIJ} We do not believe the
peak to be a narrow Kondo resonance since it is present in the
Coulomb blockade regions with even filling as well. Figure
\ref{Fig:Kondos}d shows the zero bias peak area versus gate voltage
across the Kondo resonance. The peak is maximum in each Coulomb
blockade resonance and also present in the Kondo resonance, while it
is heavily suppressed in the Coulomb blockade regions.

Figures \ref{Fig:Kondos}c and \ref{Fig:Kondos}e show another Kondo
resonance (K2) with a smaller zero bias peak. Similar behavior is
seen in this Kondo resonance: Maxima of the zero bias peak area in
the Coulomb blockade resonances and a small finite value in the
Kondo resonance region, while the zero bias peak area is fully
suppressed entering Coulomb blockade. These behaviors are
qualitatively expected for a supercurrent and could be consistent
with K2 having a lower Kondo temperature than K1. We are
unfortunately not able to turn off the superconductivity in the
leads in the setup used and more experimental work is needed to
determine the relation between the zero bias peak area and $k_B
T_K/\Delta$.

\section{Conclusion} We have succeeded in making weak links between
niobium based leads, where the link is a high quality SWCNT. At low
temperature the devices behave as quantum dots with observation of
Coulomb blockade and Kondo effect. In the case of closed dots the
onset of quasiparticle tunneling is clearly revealed. For more open
dots a zero bias peak interpreted as a supercurrent is enhanced in
Coulomb blockaded diamonds with an odd number of electron compared
to diamonds with an even number due to the presence of a Kondo
resonance.

\subsection*{Acknowledgments}
We like to thank J\o rn Bindslev Hansen and Inge Rasmussen for
assistance and use of their niobium sputtering machine. This work is
supported by the Danish Technical Research Council (The
Nanomagnetism framework program), EU-STREP Ultra-1D program and the
Nano-Science Center, University of Copenhagen, Denmark.

\end{document}